\documentstyle[preprint,eqsecnum,epsfig,aps]{revtex}

\newcommand{\newc}{\newcommand}
\newc{\be}{\begin{equation}}
\newc{\ee}{\end{equation}}
\newc{\bea}{\begin{eqnarray}}
\newc{\eea}{\end{eqnarray}}
\begin{document}
\title{Neutrinoless Double Beta Decay of $^{76}Ge$, $^{82}Se$, $^{100}Mo$ and
$^{136}Xe$ to excited $0^+$ states
}
\author{ 
F. \v Simkovic$^{1,3)}$
M. Nowak$^{2)}$, W.A. Kami\'nski$^{2)}$, 
A.A. Raduta$^{1,4)}$\
and Amand Faessler$^{1)}$
}
\address{
$^1$  Institute of Theoretical Physics, University of Tuebingen,
D--72076 Tuebingen, Germany\\
$^2$ Department of Theoretical Physics,
Maria Curie--Sklodowska University, PL-20 031 Lublin, Poland \\
$^3$Department of Nuclear Physics,  
Comenius University, Mlynsk\'a dolina F1, 
SK--842 15 Bratislava, Slovakia\\
$^4$ Institute of Physics and Nuclear Engineering, 
Bucharest, POB MG6, Romania and
Dept. of Theoretical Physics and
Mathematics, Bucharest University, POB MG11, Romania }
\date{\today}
\maketitle
\draft
\begin{abstract}
{The neutrinoless double beta decay ($0\nu\beta\beta$-decay) transition to 
the first excited $0^+$ collective final state is examined for A=76, 82, 100 
and 136 nuclei by assuming light and heavy Majorana neutrino exchange
mechanisms as well as the trilinear R-parity violating 
contributions. Realistic 
calculations of nuclear matrix elements have been performed within the 
renormalized Quasiparticle Random Phase Approximation. Transitions to the 
first  excited two-quadrupole phonon $0^+$ state 
are described within a boson expansion formalism and alternatively by using 
the operator recoupling method. We present the sensitivity parameters to 
different lepton number violating signals, which can be used in planning the
$0\nu\beta\beta$-decay experiments. The half-life of 
$0\nu\beta\beta$-decay to the first excited state  $0^+_1$  is by a factor of 
$10$ to $100$ larger than that of the  transition to the ground  state, 
$0^+_{g.s.}$. 

{PACS numbers:23.40.BW,23.40.HC}}
\end{abstract}
\section{Introduction}
\label{sec:level1}
The neutrinoless double beta decay ($0\nu\beta\beta$-decay),
which violates the total lepton number by two units,
is the most sensitive low-energy probe for physics beyond the
standard model (SM) \cite{doi83,hax84,pv,fae98,moh99}. The 
observation of the $0\nu\beta\beta$-decay
would provide unambiguous evidence that at least one of the 
neutrinos is a Majorana particle with non-zero mass \cite{sch82}. 
This conclusion is valid without specifying which from the
plethora of possible  $0\nu\beta\beta$-decay mechanisms 
triggered by exchange of neutrinos, neutralinos, gluinos, 
leptoquarks etc is the leading one. The current experimental
upper limits on the $0\nu\beta\beta$-decay half-life impose stringent
constraints, e.g.,  on  the parameters 
of Grand Unification (GUT) and supersymmetric (SUSY) extensions 
of the SM. 

There is a continuous, both experimental and theoretical, 
activity in the field of  $0\nu\beta\beta$-decay. An interesting 
issue is what are the implications of the neutrino oscillation phenomenology
to the $0\nu\beta\beta$-decay. We note that 
the results of the solar \cite{sola}, atmospheric \cite{fuku} 
and terrestrial\cite{lsnd}  neutrino experiments
provide a convincing evidence of neutrino oscillations, which require
non-vanishing masses for neutrinos as well as neutrino mixing\cite{bil}. 
The neutrino
oscillations  are sensitive to the differences of the masses squared
and cannot distinguish between Dirac and Majorana neutrinos. 
Nevertheless, if assumptions about the character 
(Dirac or Majorana neutrinos), the phases  and
 the neutrino mixing pattern are considered, one
can derive estimates for the
effective Majorana electron neutrino mass 
$<m_{\nu}>$ responsible for $0\nu\beta\beta$-decay. The current viable
analysis implies the effective neutrino mass
$<m_{\nu}>$ to be within the range 
$10^{-3} ~eV~\le~<m_{\nu}>~\le~1~eV$ \cite{bil99,haug}. The present generation 
of $0\nu\beta\beta$-decay experiments \cite{bau99,avi,priv}
achieve sensitivities
of $T_{1/2}^{0\nu} \sim 10^{24}-10^{25}$ years for different isotopes
\cite{bau99,avi,priv}. It  corresponds to 
$<m_{\nu}> \approx 0.5-1.~ eV$ \cite{si99,panfs}, i.e., 
these  $0\nu\beta\beta$-decay experiments
allow already to discriminate among various neutrino 
mixing schemes. 

The neutrino oscillations imply that perhaps we are 
close to the observation of the $0\nu\beta\beta$-decay.
This would be a major
achievement. Maybe it is enough to increase the sensitivity to $<m_{\nu}>$ 
by about an
order of magnitude, i.e., the $0\nu\beta\beta$-decay experiments
should be sensitive to half-lifes of  $10^{27}-10^{28}$ years for
the ground  to ground  transitions. We hope that these data will
stimulate new experimental activities. Ambitious plans are underway to 
push the upper constraints on lepton number violating parameters  further
down. By using several tons of enriched $^{76}Ge$, the GENIUS
experiment is expected to probe $<m_{\nu}>$ up to $10^{-2}~eV$ \cite{genius}.
The CUORE experiment intends to search for rare events with the 
help of a cryogenic
$TeO_2$ detector with high energy resolution \cite{cuore}. 
The ongoing NEMO 3 experiment,
now under construction 
in the Fr\'ejus underground laboratory, will measure up to 
10 kg of different double beta decay isotopes \cite{nemo3}. Both CUORE
and NEMO 3 have a chance to reach a sensitivity to the effective neutrino 
mass $<m_{\nu}>$ in the order of $0.1~eV$ \cite{panfs}

It is worth to examine also other possibilities to  
increase the sensitivity 
of $0\nu\beta\beta$-decay experiments.
Till now, the attention was concentrated mostly to the  $0\nu\beta\beta$-decay
transition to the ground state of the final nucleus. However, there
might be a chance
that the transitions to the excited $0^+$ and/or $2^+$ final states are
more favorable experimentally, at least for a particular mechanism
for the  $0\nu\beta\beta$-decay \cite{priv,alba}. Generally speaking, transitions
to the excited states are suppressed due to the reduced $Q_{\beta\beta}$
value. However, this restriction can be compensated by a possible 
lower background due to a 
coincidence of the $\beta$ particles with the $\gamma$ or $\gamma$'s
from the excited final state. 
The possible advantage depends on the ratio of
corresponding nuclear matrix elements to the excited and to the ground
state. If their values 
are comparable, the $0\nu\beta\beta$-decay experiment measuring
transitions to ground and excited final states could be of a
similar sensitivity.

Recently, this issue received an increasing attention from many experts
in the field. 
The $0\nu\beta\beta$-decay of $^{76}Ge$ and $^{100}Mo$
 to the first excited $2^+_1$ final state has been investigated in 
Ref. \cite{tomo} by assuming massive Majorana neutrinos   and right-handed 
weak currents. It was found that the zero neutrino double beta decay 
transition probabilities for the $0^+ \rightarrow 2^+$ decay 
are strongly suppressed 
due to higher partial waves of the emitted electrons needed 
in the  $0^+ \rightarrow 0^+_{g.s.}$transition. But in the case
of the $0\nu\beta\beta$-decay to lowest and first excited $0^+$ 
states the two emitted electrons are preferentially in a $s_{1/2}$
wave.  Therefore, this decay channel is more favored. 
Recently, the first realistic calculation for  $0^+ \rightarrow 0^+_1$
decay  has been performed for A=76 and 82 nuclei
within a higher Quasiparticle Random Phase Approximation (QRPA)
\cite{suhi}. The $0\nu\beta\beta$-decay
mechanisms mediated by light Majorana neutrinos within the
left-right symmetric model have been discussed. The conclusion 
was that the transition to excited collective $0^+$ final states 
are reduced compared with the decay to the ground state. It would be
worthwhile  
to test  this result also within other nuclear approaches.

The aim of this work is to examine the  $0\nu\beta\beta$-decay of 
$^{76}Ge$ $^{82}Se$, $^{100}Mo$ and  $^{136}Xe$ to the first
excited $0^+_1$  state. We shall discuss the mechanisms
induced by light and heavy Majorana neutrino exchange as well as
those with trilinear R-parity violation. The nuclear matrix elements will
be evaluated within the renormalized QRPA (RQRPA) \cite{toi,simn96}, which 
take into account the Pauli exclusion principle. 
We shall also consider the contributions
to  $0\nu\beta\beta$-decay coming from the momentum-dependent 
induced nucleon currents, which has been found to be significant for
ground state to ground state transitions\cite{si99}. 
We note that this contributions 
were ignored in a similar study  \cite{suhi}. The 
collective  two-quadrupole phonon state  $0^+_1$, will be  
described by two different approaches proposed 
in Refs.  \cite{grif}  and  \cite{radu91,adu,rad20}, respectively. Finally
the sensitivity parameters for a given isotope,  associated with different  
lepton number violating signals and nuclear transitions, will be calculated.
Also a discussion about possible projects of future experimental searches for
$0\nu\beta\beta$-decay to excited  final states $0^+$, will be presented.

The paper is organized as follows. In section II, 
basic formulae relevant to the  $0\nu\beta\beta$-decay 
mechanisms, are presented. In section III two approaches meant 
to calculate the transitions
to ground and excited states, are described.  
In section IV we calculate the $0\nu\beta\beta$-decay matrix elements
for A=76, 82, 100 and 136 nuclei
via RQRPA. We also determine the sensitivity of transitions to 
the collective excited  state $0^+_1$, to  lepton number violation 
associated with the Majorana neutrino mass and R-parity breaking.
The perspectives of measuring $0^+ \rightarrow 0^+_1$ decays are
analyzed. The summary and final conclusions are presented in Sec. IV.

\section{The $0\nu\beta\beta$-decay half-life}
\label{sec:level2}
The theory of light and heavy Majorana neutrino mass modes 
of  $0\nu\beta\beta$-decay 
have been reviewed, e.g., in Refs. \cite{doi83,hax84,fae98,si99}. 
The trilinear R-parity violating mode of  $0\nu\beta\beta$-decay 
has been presented in Refs. \cite{Moh86,Ver87,awf99,FKSS97}.
Without going into details of derivations, we
summarize the basic ingredients of these modes of   
$0\nu\beta\beta$-decay.

\subsection{Majorana neutrino mass mechanism}

The half-life of $0\nu\beta\beta$-decay associated with
light and heavy Majorana neutrino mass mechanism is given as
\begin{equation}
[T_{1/2}^{0\nu}]^{-1} = G_{01} 
|\frac{<m_\nu >}{m_e} M^{light}_{<m_\nu >} + 
\eta_{_{N}} M^{heavy}_{\eta_{_N}}|^2.
\label{eq:1}   
\end{equation}
Here, $G_{01}$ is the integrated kinematical factor \cite{doi83,pan96}. 
The lepton number non-conserving parameters, i.e., the
effective electron Majorana neutrino mass $<m_\nu >$ and 
$\eta^{}_{_N}$,   are given as follows:
\begin{eqnarray}
<m_\nu > ~ = ~ \sum^{3}_1~ (U^L_{ek})^2 ~ \xi_k ~ m_k, ~ ~~~~~~~
\eta_{_N} ~ = ~ \sum^{3}_1~ (U^L_{ek})^2 ~ 
~\Xi_k ~ \frac{m_p}{M_k},
\label{eq:2}   
\end{eqnarray}
with $m_p$ ($m_e$) being the proton (electron) mass. $U^L$ is the unitary
mixing matrix connecting left-handed neutrino weak eigenstates $\nu_{l L}$,
to mass eigenstates of light $\chi_{k}$ and heavy $N_{k}$ Majorana
neutrinos with masses $m_k$ ($m_k << 1$ MeV) and $M_k$ ($M_k >> 1$ GeV),
respectively. We have
\begin{equation}
\nu_{l L}
 = \sum_{k=light} ~U^L_{l k}~\chi_{kL} +
\sum_{k=heavy} ~U^L_{l k}~N_{kL}~~~~~(l = e, ~\mu~\tau).
\label{eq:3}   
\end{equation}
$\nu_k,N_k$ satisfy the Majorana condition: 
$\nu_k \xi_k = C ~{\overline{\nu}}_k^T$, 
$N_k \Xi_k = C ~{\overline{N}}_k^T$,
where C denotes the charge conjugation while $\xi$, $ \Xi$ 
are phase factors; the eigenmasses are assumed positive. 

The nuclear matrix elements associated with the
exchange of light ($M^{light}_{<m_\nu >}$) and heavy neutrinos
($M^{heavy}_{\eta_{_N}}$), including  contributions from induced
nucleon currents can be written as a sum 
of Fermi, Gamow-Teller and tensor components \cite{si99}:
\begin{equation}
{\cal M}_{\cal K} = -\frac{M^{\cal K}_F}{g^2_A} + M^{\cal K}_{GT} + M^{\cal K}_{T}
~~~({\cal K} = <m_\nu>,~\eta_{_N}).
\label{eq:4}   
\end{equation}
Here, $g_A = 1.25$. Using the second quantization formalism,  
${\cal M}_{\cal K}$ can be expressed in terms of relative coordinates
as follows:
\begin{eqnarray}
\label{eq:5}
{\cal M}_{\cal K} & = &
\sum_{J^{\pi}} \sum_{{{p n p' n' } \atop m_i m_f {\cal J}  } }
(-)^{j_{n}+j_{p'}+J+{\cal J}}(2{\cal J}+1)
\left\{\matrix{
j_p & j_n & J \cr
j_{n'} & j_{p'} & {\cal J}}\right\} \nonumber \\
&& \times
\langle p(1), p'(2);{\cal J} | f(r_{12}) \tau_1^+ \tau_2^+ 
{\cal O}_{\cal K}  (12)
f(r_{12}) |n(1) ,n'(2);{\cal J}\rangle \nonumber \\
&& \times
\langle 0_f^+ \parallel
\widetilde{[c^+_{p'}{\tilde{c}}_{n'}]_J} \parallel J^\pi m_f\rangle
\langle J^\pi m_f|J^\pi m_i\rangle
\langle J^\pi m_i \parallel [c^+_{p}{\tilde{c}}_{n}]_J \parallel
0^+_i\rangle.
\end{eqnarray}
Here, $f(r_{12})$ is the short-range correlation function  \cite{fae98}
and ${\cal O}_{\cal K}(12)$ (${\cal K} =  <m_\nu>,~\eta_{_N}$) 
represents the coordinate and spin dependent part of the
two-body $0\nu\beta\beta$-decay transition operator
\begin{equation}
{\cal O}_{\cal K} (12) = 
- \frac{H_{F}^{\cal K} (r_{12})}{g^2_A}   + 
H_{GT}^{\cal K} (r_{12}) {\bf \sigma}_{12} +
H_{T}^{\cal K} (r_{12}) {\bf S}_{12}.
\label{eq:6}
\end{equation}
Also, the following notations have been used:
\begin{eqnarray}
{\bf r}_{12} &= &{\bf r}_1-{\bf r}_2, ~~~
r_{12} = |{\bf r}_{12}|, ~~~
\hat{{\bf r}}_{12} = \frac{{\bf r}_{12}}{r_{12}},~\nonumber \\
S_{12} &=& 3({\vec{ \sigma}}_1\cdot \hat{{\bf r}}_{12})
       ({\vec{\sigma}}_2 \cdot \hat{{\bf r}}_{12})
      - \sigma_{12}, ~~~ \sigma_{12}=
{\vec{ \sigma}}_1\cdot {\vec{ \sigma}}_2.
\label{eq:7}   
\end{eqnarray}
 where ${\bf r}_1$ and ${\bf r}_2$ are coordinates of the beta decaying nucleons.
The radial part of the light and heavy neutrino exchange potentials
 $H^{<m_\nu >}_{I}(r_{12}) $ and  $H^{\eta_{_N} }_{I}(r_{12})$
($I = F, ~GT, ~T$) can be written as
\begin{eqnarray}
H^{<m_\nu >}_{I}(r_{12}) &=&\frac{2}{\pi g^2_A}
\frac{R}{r_{12}} \int_{0}^{\infty} 
\frac{\sin(qr_{12})} {q+E^m(J)- (E^i + E^f)/2} h_{I}(q^2)\,d{q}, 
\nonumber \\
H^{\eta_{_N} }_{I}(r_{12}) &=& \frac{1}{m_p m_e}\frac{2}{\pi g^2_A}
\frac{R}{r_{12}} \int_{0}^{\infty} 
{\sin(qr_{12})}h_{I}(q^2)~ q~ d{q} 
\label{eq:8}   
\end{eqnarray}
with
\begin{eqnarray}
h_{F}(q^2) &=& g^2_V (q^2) g^2_A, \nonumber \\
h_{GT}(q^2) & = & 
g^2_A (q^2) +\frac{1}{3}\frac{g^2_P (q^2) q^4}{4 m^2_p}
-\frac{2}{3}\frac{g_A (q^2) g_P (q^2) q^2}{2 m_p}
+\frac{2}{3} \frac{g^2_M ({\vec q}^{~2}) {\vec q}^{~2} }{4m^2_p}, \nonumber \\
h_{T}(q^2) & = & 
\frac{2}{3}\frac{g_A (q^2) g_P (q^2) q^2}{2 m_p}
- \frac{1}{3}\frac{g^2_P (q^2) q^4}{4 m^2_p}
+\frac{1}{3} \frac{g^2_M ({\vec q}^{~2}) {\vec q}^{~2}}{4m^2_p}.
\label{eq:9}   
\end{eqnarray}
Here, $R=r_0 A^{1/3}$ is the mean nuclear radius \cite{si99} with $r_0=1.1~fm$. 
$E^i$, $E^f$ and $E^{m}(J)$ are  the energies of the 
initial, final and intermediate nuclear states with angular momentum  $J$,
 respectively.
The momentum dependence of the vector, weak magnetism, axial-vector and pseudoscalar
formfactors ($g_V(q^2)$, $g_M(q^2)$, $g_A(q^2)$ and $g_P(q^2)$) can be found
in Ref. \cite{si99}.

We note that the overlap factor $\langle J^\pi m_f|J^\pi m_i\rangle$
and the one--body transition densities 
$<J^\pi m_i\parallel [c^+_{p}{\tilde{c}}_{n}]_J\parallel 0^+_i>$
and 
$<0_f^+\parallel\widetilde{ [c^+_{p}{\tilde{c}}_{n}]_J}\parallel J^\pi m_f> $
entering Eq. (\ref{eq:5}) must be computed in a nuclear model.

\subsection{The trilinear R-parity violating mechanisms}

The minimal supersymmetric standard model (MSSM), which is the simplest 
extension of the SM,  preserves the R-parity, i.e., 
also the total lepton number. We remind that
$R$-parity is a discrete multiplicative  symmetry
defined as $R_p=(-1)^{3B+L+2S}$, where $S, B$ and $L$
are the spin, the baryon and the lepton quantum number.
Thus $R_p=+1$ for SM particles and $R_p = -1$ for superpartners.

Generally, one can add trilinear R-parity violating terms to 
the superpotential of MSSM
\begin{eqnarray}
\label{W_rp}
W_{R_p \hspace{-0.8em}/\;\:} = \lambda_{ijk}L_i L_j E^c_k +
\lambda^{\prime}_{ijk}L_i Q_j D^c_k + \mu_j L_j H_2+
\lambda^{\prime\prime}_{ijk} U^c_i D^c_j D^c_k,
\label{eq:10}   
\end{eqnarray}
where $i,j,k$ denote generation indices.
Here $L$,  $Q$  stand for the lepton and quark doublet left-handed 
superfields while $ E^c, \  U^c,\   D^c$
for the charge conjugated
lepton, {\em up} and {\em down} quark singlet  superfields, respectively.
The terms proportional to $\lambda$ and $\lambda'$ violate the lepton number 
while
those proportional to $\lambda''$ violate the baryon number. 

The $0\nu\beta\beta$-decay can be induced by different 
trilinear R-parity violating mechanisms, partially determined by
different products of the parameters $\lambda$ and $\lambda'$ 
\cite{haug,FKSS97,pas99}.
Here, we consider those mechanisms which lead to 
the most stringent constrain on the ${\lambda'}_{111}$ parameter.
They are triggered by exchange of gluinos and neutralinos.
The corresponding Feynman diagrams can be found,  e.g., 
in Ref.  \cite{panfs,FKSS97}. 
If the masses of the SUSY particles are assumed to be of about the same
value, there is a dominance of the gluino-exchange mechanism \cite{FKSS97}. 
This conclusion is expected to be valid also for the 
$0\nu\beta\beta$-decay transitions to excited $0^+$ states. 

The $0\nu\beta\beta$-decay half-life, associated 
with  exchange of gluinos,  is \cite{fae98,FKSS97}
\begin{equation}
\big[ T_{1/2}(0^+ \rightarrow 0^+) \big]^{-1} = 
G_{01}~ |\eta_{\tilde g} + 4 \eta'_{\tilde g}|^2~ 
  |{\cal M}_{{\lambda'}_{111}}|^2.
\label{eq:11}   
\end{equation}
The effective $R_p \hspace{-1em}/\;\:$ violating parameters
$\eta_{\tilde g}$ and $\eta'_{\tilde g}$ can be expressed 
by means of the fundamental parameters of the MSSM,
as follows:
\begin{eqnarray}
\eta_{\tilde g} &=& \frac{\pi \alpha_s}{6}
\frac{\lambda^{'2}_{111}}{G_F^2 m_{\tilde d_R}^4} \frac{m_P}{m_{\tilde g}}\left[
1 + \left(\frac{m_{\tilde d_R}}{m_{\tilde u_L}}\right)^4\right],
\nonumber \\
\eta'_{\tilde g} &=& \frac{\pi \alpha_s}{12}
\frac{\lambda^{'2}_{111}}{G_F^2 m_{\tilde d_R}^4}
\frac{m_P}{m_{\tilde g}}
\left(\frac{m_{\tilde d_R}}{m_{\tilde u_L}}\right)^2.
\label{eq:12}   
\end{eqnarray}
Here, $\alpha_s = g_{3}^{2}/(4\pi)$ is the $SU(3)_c$ gauge 
coupling constant. $m_{{\tilde u}_L}$, $m_{{\tilde d}_R}$ and 
$m_{\tilde g}$ are masses of the u-squark, d-squark and 
the gluino.

At the level of hadronization, the dominant mechanism is
the pion realization of the underlying $\Delta L = 2$
quark-level $0\nu\beta\beta$-transition $d d\rightarrow u u + 2 e^-$
\cite{FKSS97}. The nuclear matrix element  ${\cal M}_{{\lambda'}_{111}}$
can be written as a sum of contributions originating from
one and two pion--exchange modes. Thus, we have \cite{FKSS97}
\begin{eqnarray}
{\cal M}_{{\lambda'}_{111}} &=& {\cal M}^{1\pi} + {\cal M}^{2\pi}
\nonumber \\
&=& \left(\frac{m_A}{m_{_p}}\right)^2
\frac{m_{_p}}{ m_e}
      \left(
\frac{4}{3} \alpha^{1\pi} ( M^{1\pi}_{GT} + M^{1\pi}_{T}) + 
\alpha^{2\pi} (M^{2\pi}_{GT} + M^{2\pi}_T) \right).
\label{eq:13}   
\end{eqnarray}
where
\begin{eqnarray}
\label{MGT}
M_{GT}^{k\pi} &=&
\langle 0^+_f|\sum_{i\neq j} \tau_i^+ \tau_j^+
{\vec \sigma}_i \cdot {\vec \sigma}_j
F_{GT}^{(k)}(m_\pi r_{\pi}) \frac{R}{r_{ij}}
|0^+_i\rangle\,,\quad\mbox{with}\quad k=1,2 \nonumber\\
M_{T}^{k\pi}&=&
\langle 0^+_f|\sum_{i\neq j} \tau_i^+ \tau_j^+
\left[3({\vec \sigma}_i\cdot \hat{{\bf r}}_{ij})
       (\vec{\sigma_j}\cdot \hat{{\bf r}}_{ij})
      - {\vec \sigma}_i\cdot {\vec \sigma}_j\right]
      F_{T}^{(k)}(m_\pi r_{\pi})\frac{R}{r_{ij}}
|0^+_i\rangle \,,
\label{eq:14}   
\end{eqnarray}
with  
\begin{eqnarray}
F_{GT}^{(1)}(x) &=&  e^{- x}, \ \ \ F_{T}^{(1)}(x) =
(3 + 3x + x^2)\frac{e^{- x}}{x^2},\\
F_{GT}^{(2)}(x) &=& (x - 2) e^{- x}, \ \ \ F_{T}^{(2)}(x) = (x + 1) e^{- x}.
\label{eq:15}
\end{eqnarray}
Here, $m_A (= 850~MeV)$ and $m_\pi$ are the mass scale of nucleon formfactor
and the mass of the pion, respectively.
Values of the structure coefficients $\alpha^{k\pi}$ (k=1,2)
are \cite{FKSS97}: $\alpha^{1\pi} = -4.4\times 10^{-2}$,   
$\alpha^{2\pi} = 2\times 10^{-1}$.   

Having in mind the forthcoming calculations within the RQRPA, it
is useful to rewrite  ${\cal M}_{{\lambda'}_{111}}$ in the form given by
Eqs. (\ref{eq:5}) and (\ref{eq:6})  
(${\cal K} = {{\lambda'}_{111}}$). One finds that the Fermi part 
of the pion--exchange potential is equal to zero 
($H^{{\lambda'}_{111}}_{F}(r_{12})=0$)
and the the Gamow-Teller and tensor parts are given by:
\begin{eqnarray}
H^{{\lambda'}_{111}}_{GT}(r_{12}) &=&
\left(\frac{m_A}{m_{_p}}\right)^2
\frac{m_{_p}}{ m_e} \left(
\frac{4}{3} \alpha^{1\pi}  F^{1\pi}_{GT}(m_\pi r_{12}) + 
\alpha^{2\pi}  F^{2\pi}_{GT}(m_\pi r_{12})  
\right),
\nonumber \\
H^{{\lambda'}_{111}}_{T}(r_{12}) &=& 
\left(\frac{m_A}{m_{_p}}\right)^2
\frac{m_{_p}}{ m_e} \left( 
\frac{4}{3} \alpha^{1\pi} F^{1\pi}_T(m_\pi r_{12})
 + \alpha^{2\pi} F^{2\pi}_T (m_\pi r_{12}) 
\right).
\label{eq:16}   
\end{eqnarray}

\section{ One--body transition densities}
\label{sec:level3}
The nuclear matrix element  
${\cal M}_{\cal K}$ (${\cal K} =  <m_\nu>,~\eta_{_N}~{\lambda'}_{111}$), 
in Eq. (\ref{eq:5}), is associated with  the $0\nu\beta\beta$-decay
 to the  ground state, $0^+_{g.s.}$, or to any of the  excited $0^+$ 
states in the final nucleus. 
Its evaluation requires the description of the
initial  ($|0^+_i\rangle$), final  ($|0^+_f\rangle$)
and the intermediate states (all in different nuclei) 
with angular momentum and parity $J^\pi$
($| J^\pi m_{i,f}\rangle$),  within a given nuclear model.
Then, the one-body transition density, entering the expression
for ${\cal M}_{\cal K}$, can be calculated and consequently 
the  chosen  matrix element is readily obtained.

The standard  QRPA (based on the quasiboson approximation)
and the RQRPA have been intensively used 
to calculate nuclear matrix elements for the double beta decay
 \cite{fae98,si99,suhi,toi,simn96,awf99,FKSS97,pan96}.
The RQRPA includes anharmonicities (the ground state is less correlated
than in the standard QRPA)  and there is no collapse of its
first solution within the
physical range of the particle-particle interaction strength. 
Within schematic models, it has been shown that by  
including  the Pauli Exclusion
Principle (PEP) in the QRPA, good agreement with the exact solution of 
the many-body
problem can be achieved even beyond the critical point of the standard QRPA
\cite{schm}. The RQRPA takes into account the PEP in an approximate way.
Nevertheless, it is enough to avoid the 
main drawback of the standard QRPA and
to reduce the sensitivity of the calculated observables to the details
of the nuclear model. The RQRPA has been used in our previous studies
of the double beta decay \cite{fae98,si99,rad20,awf99,FKSS97,wie}.
Here we apply this approach to calculate the $0\nu\beta\beta$-decay
to first excited  states $0^+$.

The final nuclei for A = 76, 82, 100 and 136 double beta decaying systems
are $^{76}$Se, $^{82}$Kr, $^{100}$Ru and $^{136}$Ba, respectively. 
The first excited $0^+_1$ state of these nuclei is believed to be  member
of the  vibrational triplet  $0^+$, $2^+$, $4^+$ . This state can  be described 
as follows
\begin{equation}
|0^+_1 \rangle = \frac{1}{\sqrt{2}} 
\{\Gamma^{1\dagger}_2 \otimes \Gamma^{1\dagger}_2\}^0
|0^+_{g.s.} \rangle, 
\label{eq:17}
\end{equation}
where $\Gamma^{1\dagger}_2$ is the creation quadrupole phonon operator.
The experimental energies of
$0^+_1$ ($E(0^+_1)$) 
and $2^+_1$ ($E(2^+_1)$) states relative to the ground state energies
are
\begin{eqnarray}
 [ E(0^+_1), E(2^+_1) ]  &=& [ 1.122, 0.559 ] ~MeV~ \rm{for}~A=76, \nonumber\\
  &=& [ 1.488, 0.777 ] ~MeV~ \rm{for}~A=82, \nonumber\\
  &=& [ 1.130, 0.540 ] ~MeV~ \rm{for}~A=100, \nonumber\\
  &=& [ 1.579, 0.818 ] ~MeV~ \rm{for}~A=136. 
\label{eq:18}
\end{eqnarray}
One notices that the energy of the $0^+_1$ excited state is about
twice the energy of the  $2^+_1$ excited state.

The nuclear states of interest are described as
charge changing (pn-RQRPA) and charge conserving (ppnn-RQRPA)
modes of the RQRPA approach. 

In the framework of the pn-RQRPA,
the $m^{th}$ excited state of the intermediate odd-odd nucleus,
with the angular momentum $J$ and the projection $M$, 
is created by applying the phonon-operator $Q^{m\dagger}_{JM^\pi}$
on the vacuum state $|0^+_{RPA}\rangle$:
\begin{equation}
|m, JM^\pi \rangle = Q^{m\dagger}_{JM^\pi}|0^+_{RPA}\rangle
 \qquad \mbox{with} \qquad
Q^{m}_{JM^\pi}|0^+_{RPA}\rangle=0.
\label{eq:19}
\end{equation}
Here $|0^+_{RPA}\rangle$ is the ground state of the initial  or 
the final nucleus and
the phonon-operator $Q^{m\dagger}_{JM^\pi}$  is defined by the ansatz:
\begin{equation}
Q^{m\dagger}_{JM^\pi} = \sum_{pn}
 \left [ X^m_{(pn, J^\pi)} A^\dagger(pn, JM)
+ Y^m_{(pn, J^\pi)}\tilde{A}(pn, JM)\right ].
\label{eq:20}
\end{equation}
$X^m_{(pn, J^\pi)}$, $Y^m_{(pn, J^\pi)}$ denotes free 
variational amplitudes, which are calculated by solving
the RQRPA equations.

The first excited $2^+$ state of the daughter nucleus is assumed
to have one quadrupole--phonon character. 
Within the ppnn-RQRPA (allowing for two 
proton and two neutron quasiparticle excitations only) this
state is defined by:
\begin{equation}
|2^+_1 \rangle = \Gamma^{1\dagger}_{2M^+}|0^+_{RPA}\rangle
 \qquad \mbox{with} \qquad
\Gamma^{1}_{2M^+}|0^+_{RPA}\rangle=0,
\label{eq:21}
\end{equation}
where
\begin{eqnarray}
\Gamma^{1\dagger}_{2M^+} &= & \sum_{p \le p'}
 \left [ R^1_{(p,p', 2^+)} A^\dagger(p,p', 2M)
+ S^1_{(p,p', 2^+)}\tilde{A}(p,p', 2M)\right ] +
\nonumber \\
&&\sum_{n \le n'}
 \left [ R^1_{(n,n', 2^+)} A^\dagger(n,n', 2M)
+ S^1_{(n,n', 2^+)}\tilde{A}(n,n', 2M)\right ]. 
\label{eq:22}
\end{eqnarray}
$ A^\dagger(\tau\tau', JM)$  and $ A^{}(\tau\tau', JM)$
($\tau =p,n$ and $\tau'=p',n'$) are the two quasi-particle creation and
annihilation operators coupled to the good
angular momentum $J$ with projection $M$ respectively, defined by:
\begin{eqnarray}
A^\dagger(\tau\tau', JM) &=& 
\frac{(1+(-1)^J\delta_{\tau \tau'})}{(1+\delta_{\tau \tau'})^{3/2}}
\sum^{}_{m_\tau , m_\tau' }
C^{J M}_{j_\tau m_\tau j_{\tau'} m_{\tau'} } 
a^\dagger_{\tau m_\tau} a^\dagger_{\tau' m_\tau'},
\nonumber\\
A^{}(\tau\tau', JM)&=&\left (A^\dagger(\tau\tau', JM)\right )^{\dagger}.
\label{eq:23}
\end{eqnarray}
The vacua defined by Eqs. (3.5) and (3.3) are in principle
different from each other. However the differences induce
corrections to the matrix elements considered, of higher order
and therefore they are neglected.
The quasiparticle creation and annihilation operators 
($a^{+}_{\tau m_\tau}$ and $a^{}_{\tau m_\tau}$, $\tau = p,n$) 
have been defined through the Bogoliubov-Valatin transformation 
\begin{equation}  
\left( \matrix{ a^{+}_{\tau m_{\tau} } \cr
 {\tilde{a}}_{\tau  m_{\tau} } 
}\right) = \left( \matrix{ 
u_{\tau} & v_{\tau} \cr 
-v_{\tau} & u_{\tau} 
}\right)
\left( \matrix{ c^{+}_{\tau m_{\tau}} \cr
{\tilde{c}}_{\tau m_{\tau}} 
}\right),
\label{eq:24}
\end{equation} 
where $c^{+}_{\tau m_\tau}$ ($c^{}_{\tau m_\tau}$)
denotes the particle creation (annihilation) operator 
acting on a single particle level with quantum numbers 
$(n_\tau, l_\tau, j_\tau )$. The parameters $u$, $v$
are occupation amplitudes and 
the tilde symbol indicates the  time-reversal operation, e.g.
$\tilde{a}_{\tau {m}_{\tau}}$ = 
$(-1)^{j_{\tau} - m_{\tau}}a^{}_{\tau -m_{\tau}}$. 

Let us now denote by ${\cal D}_{p n}$ and
${\cal D}_{\tau\tau'}$ ($\tau=p,n$)
the following expectation values:
\begin{eqnarray}
\langle 0^+_{RPA}|[ A(p n, JM),
A^\dagger(p'n', JM)
]|0^+_{RPA}\rangle &=&
\delta_{p p'}\delta_{n n'} 
{\cal D}_{p n},
\nonumber \\
\langle 0^+_{RPA}|[ A(\tau\tau', JM),
A^\dagger(\sigma\sigma', JM)
]|0^+_{RPA}\rangle &=&
(\delta_{\tau\sigma}\delta_{\tau'\sigma'} -
(-1)^{j_\tau + j_{\tau'} -J} \delta_{\tau\sigma'}\delta_{\tau'\sigma})
{\cal D}_{\tau\tau'}.
\label{eq:25}
\end{eqnarray}
Here, the exact expressions of the commutators are taken into account.
The calculation of ${\cal D}$ factors is discussed in Ref. 
\cite{simn96,zare}.

Solving the pn-RQRPA (ppnn-RQRPA) equations, one gets the
renormalized amplitudes $\overline{X}$, $\overline{Y}$
($\overline{R}$, $\overline{S}$)
with the usual normalization 
$\overline{X}\overline{X}-\overline{Y}\overline{Y}=1$
($\overline{R}\overline{R}-\overline{S}\overline{S}=1$).
They are related with $X$ ($R$) and $Y$ ($S$) amplitudes,
characterizing the standard QRPA phonon operator by:
\begin{eqnarray}
\overline{X}^m_{(pn, J^\pi)} &=& \sqrt{{\cal D}_{pn}} 
{X}^m_{(pn, J^\pi)},~~~
\overline{Y}^m_{(pn, J^\pi)} = \sqrt{{\cal D}_{pn}} 
{Y}^m_{(pn, J^\pi)}, 
\nonumber \\
\overline{R}^m_{(\tau\tau', J^\pi)} &=& \sqrt{{\cal D}_{\tau\tau'} } 
{R}^m_{(\tau\tau', J^\pi)},~~~
\overline{S}^m_{(\tau\tau', J^\pi)} = \sqrt{{\cal D}_{\tau\tau'} } 
{S}^m_{(\tau\tau', J^\pi)}.
\label{eq:26}
\end{eqnarray}

In the quasiparticle representation, the beta transition density operator
can be written as:
\begin{eqnarray}
[c^+_{p}{\tilde{c}}_{n}]_{JM} &=&
u_p v_n A^\dagger(pn, JM) + u_n v_p \tilde{A} (pn, JM) 
\nonumber\\
&+&
u_p u_n B^\dagger(pn, JM) - v_p v_n \tilde{B} (pn, JM).
\label{eq:27}
\end{eqnarray}

If we restrict our consideration to the ground state to
ground state $0\nu\beta\beta$-decay  we end
up with the following expressions for one-body densities
\cite{si99,simn96}
\begin{eqnarray}
\label{eq:28}   
<J^\pi m_i\parallel [c^+_{p}{\tilde{c}}_{n}]_J \parallel 0^+_i>
 &=& \sqrt{2J+1} 
(u_{p}^{(i)} v_{n}^{(i)} {\overline{X}}^{m_i}_{(pn, J^\pi)}
+v_{p}^{(i)} u_{n}^{(i)} {\overline{Y}}^{m_i}_{(pn, J^\pi)})
\sqrt{{\cal D}^{(i)}_{pn}},  \\
<0_f^+\parallel\widetilde{ [c^+_{p}{\tilde{c}}_{n}]_J}
\parallel J^\pi m_f> 
&=& \sqrt{2J+1} 
(v_{p}^{(f)} u_{n}^{(f)} 
{\overline{X}}^{m_f}_{(pn, J^\pi)}
+u_{p}^{(f)} v_{n}^{(f)} 
{\overline{Y}}^{m_f}_{(pn, J^\pi)})
\sqrt{{\cal D}^{(f)}_{pn}}
\label{eq:29}   
\end{eqnarray}
Here, the index i (f) indicates that the quasiparticles and the excited
states of the nucleus are defined with respect to the initial (final)
nuclear ground state $|0^+_i>$ ($|0^+_f>$). The overlap matrix 
elements entering 
Eq. (\ref{eq:5}) are explicitly given in Ref. \cite{simr88}. 

The beta transition density from the intermediate states $|J^\pi m>$
to the first excited $0^+$ state of the
daughter nucleus, which is considered to be of two-quadrupole phonon character
for the nuclei with A=76, 82, 100 and 136, can be written as:
\begin{equation}
\langle 0^+_1 | \widetilde{ [c^+_{p}{\tilde{c}}_{n}]_J}
| J^\pi m> 
= \langle 0^+_{RPA} | 
\frac{1}{\sqrt{2}} \{\Gamma_2 \otimes \Gamma_2\}^0
~\{ [c^+_{p}{\tilde{c}}_{n}]_{J}~\otimes  Q^{m\dagger}_{J^\pi} \}^0
|0^+_{RPA} \rangle \sqrt{2J+1}, 
\label{eq:30}   
\end{equation}
There are two basic approaches to calculate this
expression. We shall discuss them in the next sections.  

\subsection{The recoupling approach}

The first calculation of the two-neutrino double beta decay
($2\nu\beta\beta$-decay) 
transition to an excited $0^+$ final state, 
was presented in Ref. \cite{grif}. The 
formalism proposed was developed in the Tamm-Dancoff approximation (TDA).
It was claimed
that the contribution coming from the backgoing graphs are negligible.
The dominant contribution is obtained by calculating, through a
recoupling procedure, 
the scalar product of two
pairs of  proton-neutron quasiparticle creation operators,  originating
from the beta transition 
$\widetilde{[c^+_p{\tilde{c}}_n]_{JM}}$  and the phonon 
 $Q^{m\dagger}_{JM^\pi}$ operators.
\begin{eqnarray}
\{A^\dagger (pn, J) \otimes {A}^\dagger (p'n', J)\}^0
&=& - \sum_{J'} (-1)^{j_n + j_{p'} + J + J'} ~
\frac{(\delta_{pp'}(-)^{J'}+1)}{(1+\delta_{pp'})^{1/2}}~
\frac{(\delta_{nn'}(-)^{J'}+1)}{(1+\delta_{nn'})^{1/2}}
\times
\nonumber\\
&&(2J'+1)^{1/2}
\left\{\begin{array}{ccc} j_n & j_p & J\\
j_{p'} & j_{n'} & J' \end{array}\right\} 
\{A^\dagger (pp', J') \otimes {A}^\dagger (nn', J')\}^0
\label{eq:31}   
\end{eqnarray}
Henceforth we shall denote this approach as recoupling 
method (RCM).

By using  Eq. (\ref{eq:30}),  the beta transition 
matrix element takes the form
\begin{eqnarray}
&&\langle 0^+_1 | \widetilde{ [c^+_{p}{\tilde{c}}_{n}]_{JM}}
| J^\pi M, m_f> 
= \sqrt{10} \sum_{p' n'} 
(1+\delta_{p p'})^{1/2} (1+\delta_{n n'})^{1/2}
\left( \frac{ {\cal D}^{(f)}_{pp'}
{\cal D}^{(f)}_{nn'} } {{\cal D}^{(f)}_{pn} } \right)^{1/2} 
\\
&&\times \left\{ \begin{array}{ccc} j_n & j_p & J\\
j_{p'} & j_{n'} & 2 \end{array}\right\} 
 ( u_{p}^{(f)} v_{n}^{(f)} {\overline{X}}^{m_f}_{(p'n', J^\pi)}
{\overline{R}}^{m_f}_{(pp', 2^+)} {\overline{R}}^{m_f}_{(nn', 2^+)}
- v_{p}^{(f)} u_{n}^{(f)} {\overline{Y}}^{m_f}_{(p'n', J^\pi)}
{\overline{S}}^{m_f}_{(pp', 2^+)} {\overline{S}}^{m_f}_{(nn', 2^+)} ).
\nonumber 
\end{eqnarray}
\label{eq:32}
Obviously, in the above expression, the full expression of the RPA phonon
operator was used. In comparing
this transition density to $0^+_1$  with that one 
leading to the ground state, we find two important 
differences. First, the dominant contribution in Eq.(3.16) is a product
of three forward-going amplitudes. This fact implies that
the transition amplitude is not expected to be very sensitive
to the nuclear ground state correlations. Second,
the leading term in Eq. (3.16) is multiplied by the   
factor $ u_{p}^{(f)} v_{n}^{(f)}$ (i.e., ``$\beta^-$'' like)
while the leading term of beta ground state transition 
in Eq. (\ref{eq:29}) contains the factor
$ v_{p}^{(f)} u_{n}^{(f)}$ (i.e., ``$\beta^+$'' like).

The drawback of this approach is that the transition density, 
in Eq. (3.16),  contains significant unphysical
contributions. To clarify this point we transform the
second part of the r.h.s. of Eq. (3.14), which up to a
multiplicative constant should represent 
the  excited state $0^+_1$  in the (A,Z+2) nucleus.
From the transition operator written in quasiparticle representation we
keep, for illustration, the operator
 $A^\dagger(pn, J)$, which is further expressed in terms of the pnQRPA
bosons. The final result is:
\begin{eqnarray}
\{A^\dagger(pn, J) \otimes Q^{m\dagger}_{J^\pi} \}^0 
|0^+_{RPA_f} \rangle 
&=&
 \sum_{m'} [  X^{m'}_{(pn, J^\pi)} 
\{Q^{m'\dagger}_{J^\pi} \otimes Q^{m\dagger}_{J^\pi} \}^0 
|0^+_{RPA_f} \rangle 
\nonumber\\
&&+ \frac{1}{\sqrt{2J+1}}Y^{m}_{(pn, J^\pi)}
|0^+_{RPA_f} \rangle ]
\label{eq:33}   
\end{eqnarray}
The second term in the above equation, is obtained by using 
the commutator algebra for  $Q^{m\dagger}_{J^\pi}$ and its
hermitian conjugate operator, and then
Eq. (\ref{eq:19}). From Eq. (\ref{eq:33}) it follows that within
the RCM  procedure we have produced a linear combination of a state associated
to the (A,Z) nucleus and the ground state characterizing the (A,Z+2) nucleus.
We note that
the desired $0^+_1$ excited state of 
(A,Z+2) nucleus is missing. The RCM  does not allow to 
eliminate the admixture of these states, which due to the recoupling
procedure are related to the  $0^+_1$ excited state. Moreover, it is
worth noting that the component in the (A,Z+2) nucleus is proportional
to the Y-amplitude, as prescribed by the method presented in the
next subsection, and not to the 
forward going amplitude of the proton-neutron dipole phonon as
suggested by the RCM approach. 
Thus, the validity of this recoupling procedure is questionable in the 
framework of the QRPA.

The RCM has been modified by introducing a multiple commutator method 
(MCM) 
and applied to calculate  different
lepton-number conserving modes of the
 double beta decay \cite{sua}.
This version of the RCM has been also used for
describing the  $0\nu\beta\beta$-decay  
to excited collective $0^+$ states \cite{suhi}.

\subsection{The boson expansion approach}

The pioneering approach to study the double beta decay
 to excited states of the final nucleus was 
proposed in Ref. \cite{radu91,adu}. It is the so called boson expansion
method (BEM). Applications of the BEM approach to study
the single beta and $2\nu\beta\beta$-decay 
to the first excited quadrapole state ($2^+_1$) and the
two--quadrapole--phonon states ($0^+_{2-ph}$, $2^+_{2-ph}$)
of even-even isotopes were presented  in Refs. \cite{radu91,adu}.
Recently, the renormalized version of the BEM was applied 
to the transition $^{82}Se \rightarrow ^{82}Kr$ \cite{rad20}.
The new version has the virtue of exploiting the complementary features
of the BEM and RQRPA methods. As a matter of fact this improved
version of BEM is adopted in the present paper.

Within the BEM approach, the operators involved in the r.h.s. of
Eq. (\ref{eq:27}) are written as polynomials of the
RPA bosons \cite{radu91,adu}, so that the mutual commutation
relations are consistently preserved by the boson mapping.
We shall follow this procedure with some simplifications,
which do not influence the final form of
the one--body transition density.

By exploiting the fact that 
$Q^{m}_{J^\pi} |0^+_{RPA_f} \rangle = 0$, we introduce a
commutator in the expression for the one-body transition 
operator to the $0^+_1$ state. In addition, we evaluate this
commutator by satisfying exact commutation relations.
Thus we obtain
\begin{eqnarray}
\langle 0^+_1 | \widetilde{ [c^+_{p}{\tilde{c}}_{n}]_{JM}}
| J^\pi M, m_f> = (-1)^{J-M}
\langle 0^+_1 | 
[ [c^+_p{\tilde{c}}_n]_{J-M}, Q^{\dagger m}_{J^\pi} ] |0^+_{RPA_f} \rangle 
\nonumber \\
=  ( v_{p}^{(f)} u_{n}^{(f)} {{X}}^{m_f}_{(pn, J^\pi)}
+ u_{p}^{(f)} v_{n}^{(f)} {{Y}}^{m_f}_{(pn, J^\pi)} )
\sum_{\tau = p,n} 
\frac{\langle 0^+_1 | B^\dagger (\tau\tau , 00) |0^+_{RPA_f} \rangle}
{\sqrt(2j_\tau +1)}.
\label{eq:34}   
\end{eqnarray}
We omitted the terms  $A^\dagger (\tau\tau , 00)$ and
$A (\tau\tau , 00)$, since in the boson expansion formalism they
consist of
terms comprising  products of odd numbers of phonon operators 
$Q^{m\dagger}_{JM^\pi}$, $Q^{m}_{JM^\pi}$, 
$\Gamma^{1\dagger}_{2M^+}$ and $\Gamma^{1}_{2M^+}$, and consequently 
do not contribute to the above matrix element.

We proceed by performing the boson expansion of the operator 
 $B^\dagger (\tau\tau , 00)$ with the result
\begin{eqnarray}
B^\dagger (\tau\tau , 00) &=& 
{\cal B}^{20}_{11}(\tau\tau)
\{\Gamma^{1\dagger}_{2^+}\otimes \Gamma^{1\dagger}_{2^+}\}^0 
~~~~~~~~~~
\nonumber\\
&+&{\cal B}^{02}_{11}(\tau\tau)
\{\Gamma^{1}_{2^+}\otimes \Gamma^{1}_{2^+}\}^0 +
{\cal B}^{11}_{11}(\tau\tau)
\{\Gamma^{1\dagger}_{2^+}\otimes \Gamma^{1}_{2^+}\}^0.
\label{eq:35}   
\end{eqnarray}
The upper indices, accompanying the expansion coefficients, indicate the
number of the creation and annihilation phonon operators
involved in 
the given terms while the lower indices suggest that the phonon
operators correspond to the first root of the ppnn-QRPA equations.
We note that relevant to the problem studied  here is the coefficient 
${\cal B}^{20}_{11}(\tau\tau)$, which can be determined
by the following procedure. Commuting Eq. (\ref{eq:35}) 
twice with $\Gamma^{1}_{2^+}$ and then taking the expectation value 
of the result in the boson vacuum, one obtains
\begin{eqnarray} 
{\cal B}^{20}_{11}(\tau\tau) &=&
\frac{1}{2}\sum_M C^{00}_{2M~2-M} 
\langle 0| 
[ \Gamma^{1}_{2M}, [\Gamma^{1}_{2-M}, B^\dagger (\tau\tau , 00)]]
|0 \rangle
\nonumber\\
&=& - \sqrt{\frac{5}{2j_\tau+1}} 
( \sum_{\tau' (\tau < \tau')} {\overline{R}}^{1}_{(\tau\tau', 2^+)} 
{\overline{S}}^{1}_{(\tau\tau', 2^+)} 
+ \sum_{\tau' (\tau' < \tau)} {\overline{R}}^{1}_{(\tau'\tau, 2^+)} 
{\overline{S}}^{1}_{(\tau'\tau, 2^+)} ).
\label{eq:36}   
\end{eqnarray} 
In the above equations, all commutators are exactly evaluated
except for the last one for which the renormalized quasiboson approximation
is used \cite{radu91,adu,rad20}.

Now, by a  straightforward calculation, one arrives at the final expression for the
one-body transition density leading to the final excited
$0^+_1$ two phonon state:
\begin{eqnarray}
\langle 0^+_1 | \widetilde{ [c^+_{p}{\tilde{c}}_{n}]_{JM}}
| J^\pi M, m_f> 
&=& 
 ( v_{p}^{(f)} u_{n}^{(f)} {\overline{X}}^{m_f}_{(pn, J^\pi)}
+ u_{p}^{(f)} v_{n}^{(f)} {\overline{Y}}^{m_f}_{(pn, J^\pi)} )
\left( {\cal D}_{pn} \right)^{-1/2}\nonumber\\ 
&&\xi(p,p',n,n'),
\label{eq:37}   
\end{eqnarray}
where
\begin{eqnarray}
\xi(p,p',n,n') &=&\sqrt{10} [~ \frac{1}{2j_n+1} 
( \sum_{n' (n < n')} {\overline{R}}^{1}_{(nn', 2^+)} 
{\overline{S}}^{1}_{(nn', 2^+)} + 
{\overline{R}}^{1}_{(n'n, 2^+)} {\overline{S}}^{1}_{(n'n, 2^+)} 
)\nonumber \\
&&  \frac{1}{2j_p+1} 
( \sum_{p' (p < p')} {\overline{R}}^{1}_{(pp', 2^+)} 
{\overline{S}}^{1}_{(pp', 2^+)} + 
{\overline{R}}^{1}_{(p'p, 2^+)} 
{\overline{S}}^{1}_{(p'p, 2^+)} ) ~].
\label{eq:38}   
\end{eqnarray}
It is worthwhile to notice that if we replace the factor 
$\xi(p,p',n,n')$ with unity and consider 
small ground state correlations, i.e. 
${\cal D}_{pn } \simeq 1$, 
we obtain the ground state transition
density from Eq. (\ref{eq:29}).
We note that the transition density to the ground state is 
proportional to  $({\cal D}_{pn})^{1/2}$, i.e., it
is suppressed by anharmonic effects, while the transition
density to the excited $0^+_1$ state is 
proportional to  $1/({\cal D}_{pn})^{1/2}$, i.e., it is
enhanced by large ground state correlations. 

We remark that the BEM expression given in Eq. (\ref{eq:37}) differs
considerably from the  the RCM expression given 
in Eq. (3.16). We see that the BEM transition
density in  Eq. (\ref{eq:37})
consists of products of forward and backward going variational
amplitudes of the ppnn-RQRPA. It means that the final
result exhibits sensitivity to the particle-particle
interaction of the nuclear Hamiltonian. In addition, 
the BEM transition amplitude
is  a ``$\beta^+$'' like amplitude since the leading term is
proportional to $u^f_nv^f_p$. This implies a possible 
strong dependence on the ground state correlations. The fact that
the RCM and the MCM approaches differ considerably from  
the BEM procedure, when  the final state
is of multiple phonon character, was noticed already in Ref. \cite{adu}.
The BEM avoids  the operators recoupling  
and therefore the problem concerning the unphysical
RCM contributions does not appear.

\section{Numerical Results and Discussions}
\label{sec:level4}
The formalism described in the previous sections were 
applied to the transitions $^{76}Ge \rightarrow {^{76}Se}$,
$^{82}Se \rightarrow ^{82}Kr$,  $^{100}Mo \rightarrow ^{100}Ru$
and $^{136}Xe \rightarrow ^{136}Ba$. The pn-RQRPA and the
ppnn-RQRPA calculations have been performed
for the same sets  of basis-states  as in Ref. \cite{si99},
which are identical for protons and neutrons. The single
particle energies were  obtained by using a  Coulomb corrected Woods Saxon
potential.  The realistic interaction employed 
is the Brueckner G-matrix of the Bonn one--boson exchange potential.
The truncation of the single-particle space requires a renormalization 
of two--body matrix elements. The scaling of the pairing 
strength in the BCS calculation was adjusted to fit 
the empirical pairing gaps according to Ref. \cite{cheoun}.
In the RQRPA calculations, the particle--particle and 
particle--hole channels of the G-matrix 
interaction are renormalized 
by multiplying them with the parameters $g^{}_{pp}$ and $g^{}_{ph}$,
which, in principle, should be close to unity. Our adopted value
for $g_{ph}$ was  
$g_{\text{ph}} = 0.8$, as in our previous calculations \cite{fae98,si99}.
We shall present the relevant nuclear matrix elements for 
$g_{pp}=1$. Nevertheless, their sensitivity to $g_{pp}$ within the
interval 0.80-1.20, which can be regarded as physical, will be
discussed. We note that in our calculations the 
pn-RQRPA and ppnn-RQRPA channels are coupled through the 
equation for the renormalization factors D \cite{simn96}. Our 
numerical analysis shows that the quadrupole QRPA energies
of the daughter nucleus are independent of $g_{pp}$.

\begin{figure}
\vbox{
\centerline{\epsfig{file=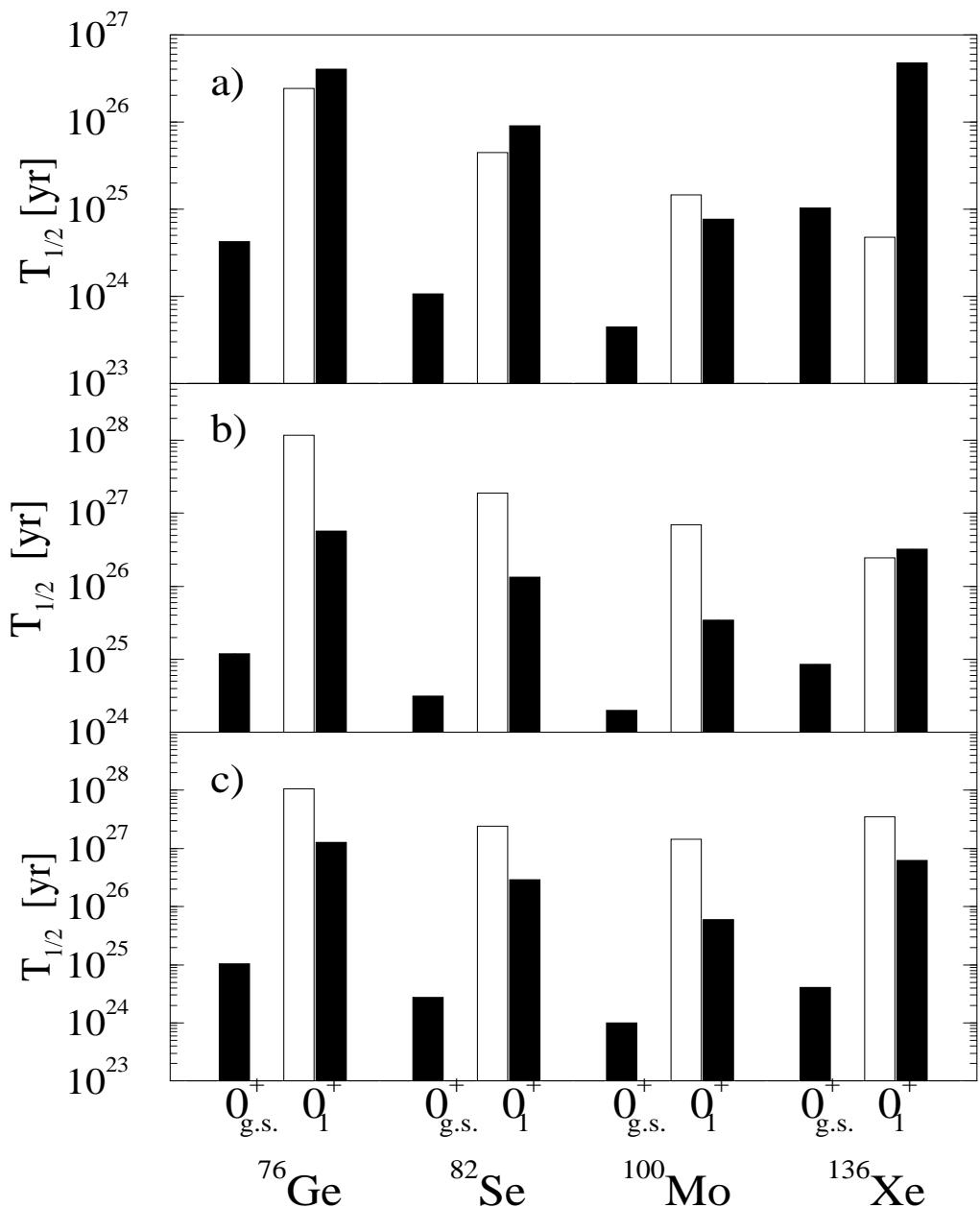,height=17.cm,width=18.cm} }
}
\caption{Calculated half-lifes of the $0\nu\beta\beta$-decay of $^{76}Ge$,
$^{76}Se$, $^{100}Mo$ and $^{136}Xe$ for transitions to the 
ground $0^+_{g.s}$ and $0^+_{1}$ excited states of the final nuclei
assuming $<m_\nu >_{ee} = 1 eV$ (a), $\eta_{_N} = 10^{-7}$ (b) 
and $\lambda'_{111} = 10^{-4}$ (c). 
The black bars  correspond to results describing ground state to ground state
transitions as well as the results obtained for the transitions to the
$0^+_1$ excited state within the boson expansion method (BEM). The open bars
denote results obtained for transitions to the $0^+_1$ state via
the recoupling method (RCM). 
}
\label{fig.1}
\end{figure}

The results of our calculations are summarized in
Tables \ref{table.1}, \ref{table.2}, \ref{table.4} 
and in Fig. \ref{fig.1}.
In Table \ref{table.1} the dimensionless
nuclear matrix elements of light and 
heavy neutrino exchange modes of $0\nu\beta\beta$-decay 
of $^{76}$Ge, $^{82}$Se, $^{100}$Mo and $^{136}$Xe, are
presented both for transitions to the ground and 
excited states. The displayed $0\nu\beta\beta$-decay 
matrix elements to the first excited states $0^+_1$, 
were obtained within the RCM and BEM approaches.
The particular contributions to the full
matrix elements coming from Fermi, Gamow-Teller 
and tensor terms in Eqs. (\ref{eq:4})  are shown as well.
The modifications coming from induced nucleon currents
are included in the Gamow-Teller and tensor components \cite{si99}.
We find that the tensor contribution plays an important
role when the mechanism is mediated by heavy neutrinos
and tends to cancel the contributions by Fermi and
Gamow-Teller transition amplitudes. 
By glancing at Table \ref{table.1} we 
find that the nuclear matrix elements involving the first excited 
$0^+_1$ state are suppressed in comparison with those associated
to transitions to the
ground state. In the case of BEM calculations of
$M^{light}_{<m_\nu >}$,  the suppression factor is about
2.8, 2.8, 1.8 and 1.5 for A=76, 82, 100 and 136, respectively.  
The RCM values are close to the BEM ones  for A=76, 82 and 100 nuclei. 
One notices an anomaly 
in the case of the A=136 system, where the RCM transition to the excited
state is by a factor of 6.7 stronger than that to the ground state.
It could be connected with the fact 
that $^{136}$Xe is a closed shell nucleus for neutrons (N=82)
and therefore the unphysical contributions to this transition
in the RCM approach might be larger. 
We note that it is not possible to compare directly    nuclear 
matrix elements $M^{light}_{<m_\nu >}$ in Table \ref{table.1}, 
with those calculated
in Ref. \cite{suhi} for A=76 and 82, since Ref. \cite{suhi}
does not include contributions
from the induced currents. Nevertheless, we note that the ratio
of the nuclear matrix elements of the transition to ground and excited states
is equal to about 3, in Ref. \cite{suhi}. This value is in good agreement 
with results of this article, despite the fact that the two formalisms 
differ from each other in many aspects.
 
From  Table \ref{table.1} it follows 
that the $0\nu\beta\beta$-decay, mediated by heavy neutrinos to 
excited final states
for A = 76, 82 and 100 
are weaker than those associated with ground state to ground state
transitions, by about a factor of 1.3-2.0  ( 8-9 )  in the BEM (RCM).
Here, the difference between the BEM and RCM predictions is
more significant than for the light Majorana neutrino
exchange mechanism. We note also that for the A=136 system, 
the RCM value of $M^{heavy}_{\eta_{_N}}$ is comparable with
 the BEM one, i.e., within the RCM, the behavior of this nucleus is
different from that of the remaining nuclei.

In Table \ref{table.2}, the nuclear matrix elements associated
with the trilinear R-parity violating mode of $0\nu\beta\beta$-decay
are displayed. Both, the one pion- and two pion--exchange 
Gamow-Teller and tensor contributions to $M^{}_{{\lambda'}_{111}}$
are shown. In Ref. \cite{awf99,FKSS97}, it was shown that there is a 
dominance of the two pion--exchange mode for the $0\nu\beta\beta$-decay 
transitions connecting the initial and final  
ground states,  due to a larger structure
coefficients $\alpha^{2\pi}$ and because of strong 
mutual cancelation
of the one-pion  exchange Gamow-Teller and tensor contributions.
We see that the second reason does not hold in the case
of transitions to $0^+_1$ excited states. We have found that
the one-pion mode plays a more 
 important role for this transition, giving a significant  
contribution to  $M^{}_{{\lambda'}_{111}}$. By comparing the 
values of nuclear matrix elements for ground and excited
state transitions, we see that the second one is reduced
by factor of 2.7-3.4 within the BEM. The RCM values are 
considerably
smaller for A=76, 82 and 100 systems. A different situation
is again found for 
the $0\nu\beta\beta$-decay of $^{136}Xe$ where the
RCM value is close to the BEM result.

One purpose of our study is also the sensitivity of
results for $M^{light}_{<m_\nu >}$, 
$M^{heavy}_{\eta_{_N}}$ and $M^{}_{{\lambda'}_{111}}$
to the details of the nuclear model. We have examined
the $0\nu\beta\beta$-transition matrix elements as function of 
the renormalization factor for the strength of the
particle--particle interaction, $g_{pp}$, 
considered in the physical interval (see Table \ref{table.3}).
We see that the BEM values for the transition to excited $0^+_1$
states exhibit a very similar dependence on $g_{pp}$  
as those for the transitions to the ground state. On the other hand,
the RCM values are insensitive to changes of $g_{pp}$.
Thus, our expectations from the previous section, hinging on
the forms of the BEM and RCM one-body transition
densities, have been confirmed.
We note that a similar behavior has been found also for other
nuclear systems.

As it was already mentioned in the introduction there is
additional suppression of the $0\nu\beta\beta$-decay to excited
$0^+$ final states coming from the smaller kinematical factor
$G_{01}$ [see Eq. (\ref{eq:1})]. 
The values of $G_{01}$  are given in Table \ref{table.4}.  
One finds that the ratio $G_{01}(0^+_{g.s.})/G_{01}(0^+_1)$ 
is about 12, 11, 5.2 and 21 for A=76, 82, 100 and 136
systems, respectively. By this factor are the 
corresponding half-lifes to excited $0^+$ final state larger.

For a given nuclear isotope
the characteristics of the $0\nu\beta\beta$-decay 
refer  both to the nuclear matrix element and the kinematical
factor. For a chosen isotope, it is worthwhile to introduce 
sensitivity parameters  with respect to different lepton number 
violating parameters. 
Large numerical values of these parameters may define
those transitions and isotopes which are the most promising
candidates for a lepton number violating
signal, in the $0\nu\beta\beta$-decay.
These parameters are defined as follows \cite{si99,panfs}:
\begin{eqnarray}
\zeta_{<m_\nu >} (Y) & = &
10^{7}~ |{\cal M}^{}_{<m_\nu >}|~ 
\sqrt{{G_{01}}~ {year}},\nonumber\\
\zeta_{\eta_{_N}} (Y)  &=&  
10^{6}~ |{\cal M}^{}_{\eta_{_N}}|~ \sqrt{{G_{01}}~{year}},
\nonumber \\
\zeta_{\lambda_{111}'} (Y) &=& 
10^{5}~ |{\cal M}^{}_{\lambda_{111}'}|~ 
\sqrt{{G_{01}}~{year}}.
\end{eqnarray}
We  listed the sensitivity parameters in  Table \ref{table.4}. 
By a glance, one finds that within
BEM the largest sensitivity parameters 
for the transitions to the $0^+_1$ state are associated with the
A=100 system followed by the A=82 system . The smallest ones are 
of the A=76 and A=136 systems, i.e., these 
transitions  are less favorable for an experimental study. Naturally,
there are also additional aspects which  experimentalists 
have to take into account in planning a search for the 
 $0\nu\beta\beta$-decay.

In Table \ref{table.4}, we present also theoretical half-lifes
by assuming $<m_\nu > = 1eV$,  $\eta_{_N} = 10^{-7}$ and 
${\acute{\lambda}}_{111} = 10^{-4}$. One finds that
in order to get a limit on  $<m_\nu > $
of 1 eV, by measuring the transition to the excited $0^+_1$ final
state, the experimentalists should reach the level of
about $10^{28}$ years for the half-life. By comparing  the theoretical 
values of $T^{0\nu}_{1/2}$ for the decay to ground state, with the
predictions for the transition to the excited state $0^+_1$, both yielded 
by BEM,
 we note that the second ones are larger by about 1-2 orders of
magnitude. This situation is shown also in Fig. \ref{fig.1}.
The question, for experimentalists, is whether the coincidence
between the deexcitation $\gamma$ and the emitted electrons allows
to reduce the background in the $0\nu\beta\beta$-decay 
experiment to a sufficient extent so that the constraints on
lepton number violating parameters deduced from the
transition to the $0^+$ excited state, can compete with those
associated with transitions to the ground state. It might be that 
in the case of the $0\nu\beta\beta$-decay of $^{100}Mo$ to 
excited $0^+_1$ state triggered by
light or heavy Majorana neutrino exchange mechanisms,
the suppression of the half-life by a factor of 17 relative to the
transition to ground state  is compensated by 
diminishing the background events. 

The expected improved experimental upper 
limits on the $0\nu\beta\beta$-decay half-life ${T^{0\nu -exp}_{1/2}}$,
imply more stringent limits on 
lepton number violating parameters $<m_\nu >$, $\eta_{_N}$ and 
${\lambda_{111}'}$. By using the sensitivity parameters
$\zeta's$ given in Table \ref{table.4}, they can be deduced in a 
straightforward way as follows:
\begin{eqnarray}
\frac{<m_\nu >}{m_e} &\leq& \frac{10^{-5}}{\zeta_{<m_\nu >}}
\sqrt{\frac{10^{24}~years}{T^{0\nu -exp}_{1/2}}}, ~~~~~~~
\eta_{_N} \leq \frac{10^{-6}}{\zeta_{\eta_{_N}}}
\sqrt{\frac{10^{24}~years}{T^{0\nu -exp}_{1/2}}}, \nonumber \\
(\lambda_{111}' )^2 &\leq&
\kappa^2~\left( \frac{m_{\tilde{q}}}{100~GeV}\right)^4~
\left(\frac{m_{\tilde{g}}}{100~GeV}\right)
\frac{10^{-7}}{\zeta_{\lambda_{111}'}}
\sqrt{\frac{10^{24}~years}{T^{0\nu -exp}_{1/2}}}
\end{eqnarray}
with $\kappa = 1.8$ \cite{fae98}. One finds that in order
to push down the upper constraint on $<m_\nu >$ below $0.1~eV$ 
in the $0\nu\beta\beta$-decay
experiment to excited $0^+_1$ final state, one
has to measure the half-life of 
 $4.02\times 10^{28}$,  $8.96\times 10^{27}$,  
$7.59\times 10^{26}$   and $4.77\times 10^{28}$ years for
A=76, 82, 100, 136 isotopes, respectively. Best present limits 
on this type of decay are on the level of $10^{21}$-$10^{22}$ 
years (see review \cite{alba}). But we would like to mention
that some progress in measuring transitions to excited states is 
expected in the future.
For example, experiment MAJORANA with 500 kg of $^{76}Ge$
plan to have sensitivity of $10^{28}$ years \cite{priv}.

\section{Summary and Conclusions}
\label{sec:level5}

The $0\nu\beta\beta$-decay is a sensitive tool to study
lepton number violating  mechanisms, which are associated
with the Majorana neutrino mass and R-parity violating
supersymmetry. Neutrino oscillations indicate that we 
may be close to the observation of this exotic rare 
process. This stimulates both  experimental and theoretical
studies, which can be helpful to limit the effective
Majorana neutrino mass $<m_\nu >$ below the 0.1 eV level.
It is an open question whether the future
$0\nu\beta\beta$-decay experiments measuring
transitions to the excited final states can
be of comparable sensitivity to different lepton
number violating parameters as 
the ground  to ground  transitions. 
Experimental studies of transitions 
to an excited $0^+_1$ final state allows to reduce the background
by  gamma-electron coincidences.
Drawbacks are lower Q values and possibly suppressed
nuclear matrix elements. The theoretical studies
of the corresponding nuclear transitions
are of great interest.

We evaluated $0\nu\beta\beta$-decay nuclear matrix elements 
for transitions to  first excited $0^+_1$ final states
for A=76, 82, 100 and 136 nuclei. The calculations
have been performed within two known approaches, 
the boson expansion method (BEM) and 
recoupling two pairs of quasiparticle operators 
(RCM). The results of these two types of calculations differ from
each other considerably especially
in the case of exchange of heavy particles for A=76, 82 and 100
systems. We indicated the drawbacks of the second approach.
We also found  anomalous behaviors of the RCM results for A=136 
nuclei. The resulting matrix elements are summarized in 
Table \ref{table.1} and Table \ref{table.2}. 
The suppression of the  decay matrix elements to $0^+_1$
in comparison with the decay to $0^+_{g.s.}$ , depends on  
the  isotope and  $0\nu\beta\beta$-decay
mechanism. An average suppression factor of about 
2-3 is predicted by the BEM. Contrary to the RCM results, the 
BEM ones are significantly
depending on the strength of the two-body interaction.

Further, we have calculated sensitivity parameters to different
signals of lepton number violation associated with
transitions to excited final states. We  compared
them with the ground  to ground 
transitions. We have found the largest sensitivity
to these parameters   in the  A=100 nuclear system. 
By comparing with the  decay to $0^+_{g.s.}$ 
we find a suppression by a factor of about 4.1
for Majorana neutrino--exchange mechanisms. It means
that the corresponding theoretical half-life is larger
 than that associated to the transition to $0^+_{g.s.}$,
by factor of 17. In order to reach the sensitivity to a
neutrino mass $<m_\nu > \approx 0.1$ eV 
in the $0\nu\beta\beta$-decay
 to an excited $0^+$  state, it is necessary
to measure half-lifes of at least $8\times 10^{26}$ years 
(as predicted by BEM). Perhaps,
this limit might be expected to be reached by the
$0\nu\beta\beta$-decay $^{100}Mo$ experiment in the near  future.

This work was supported in part by the Polish-German Agreement on
Science and Technology, grant no. 12N/99 and Pol98/050
and the State Committee for
Scientific Researches (Poland), contract no. 2P03B00516.
One of us, F.\v S wants to thank the Department  of
Theoretical Physics Maria Curie--Sklodowska University for their kind 
hospitality during his stay in Lublin and the Deutsche 
Forschungsgemeinschaft (436 SLK 17/298) for financing the stay 
in Tuebingen. A.R. wants to thank the International Buro Bonn 
(RUM-040-97) for supporting his stay in Tuebingen.


\begin{table}[t]
\caption{Nuclear matrix elements of light and heavy Majorana neutrino 
exchange 
modes for the
 $0\nu\beta\beta$-decay in A=76, 82, 100 and 136 nuclei. Both transitions
to the ground state ($0^+_{g.s.}$) and the first $0^+_1$ excited states
(which is assumed to be a two-phonon state) 
of the final nucleus are considered. The calculations have been performed 
within renormalized QRPA with the help of the recoupling  (RCM)
and the 
boson expansion (BEM) approaches for the
 evaluation of the one-body transition
density. (f.s. means ``final state'' in the table).
}
\label{table.1}
\begin{tabular}{lccccccccccc}
 & & & \multicolumn{4}{c}{Light neutrino exch. mech.} & & 
\multicolumn{4}{c}{Heavy neutrino exch. mech.}\\ \cline{4-7} \cline{9-12}
 A & f.s. & meth. &
$M^{<m_\nu >}_F$ & $M^{<m_\nu >}_{GT}$ & $M^{<m_\nu >}_T$ & 
${\cal M}_{<m_\nu >}$ &  &
$M^{\eta_{_N}}_F$ & $M^{\eta_{_N}}_{GT}$ & $M^{\eta_{_N}}_T$ & 
${\cal M}_{\eta_{_N}}$ \\ \hline
76 & $0^+_{g.s.}$ &       & 
 -1.26   &  2.18    & -0.190   & 2.80    &  &
 -37.4   &  28.7    & -20.1    &   32.6     \\
   & $0^+_1$      & RCM  & 
 -0.570  &  0.933   & -0.022  & 1.28     &  &
 -4.00   &  1.42    & -0.39   & 3.59       \\
   & $0^+_1$      & BEM  &
 -0.371  & 0.796    & -0.039  & 0.994    &  &
 -12.0  & 9.62     & -1.06   & 16.3       \\ 
82 & $0^+_{g.s.}$ &       & 
 -1.15   & 2.07    & -0.172    &  2.64   &  &
 -34.4   & 25.4    & -17.4    &  30.0      \\
   & $0^+_1$      & RCM  & 
 -0.617   &  0.971   & -0.024    &  1.34   &  &
 -4.11    &  1.40    & -0.014    &  4.02      \\
   & $0^+_1$      & BEM  &
 -0.342   &  0.762   & -0.033    &  0.947   &  &
 -11.2    &  8.61   & -0.498    &  15.2      \\ 
100 & $0^+_{g.s.}$ &       & 
  -1.28  & 2.62    & -0.230    &  3.21   &  &
  -44.3  & 34.2    & -32.9     &  29.7      \\
   & $0^+_1$      & RCM  & 
   -0.305 &  1.059    &  0.016    & 1.27    &  &
   -2.98  &  1.21    &  0.483    & 3.60    \\ 
   & $0^+_1$      & BEM  &
  -0.397  & 1.52    & -0.008     & 1.76    &  &
  -14.1  &  11.1    &  -3.90    & 16.2       \\
136 & $0^+_{g.s.}$ &       & 
 -0.504   & 0.496    & -0.161    &  0.66   &  &
 -21.7    &  16.8    & -16.6     &  14.1      \\
   & $0^+_1$      & RCM  & 
 -1.66    & 3.40  & -0.038   & 4.42  &    &
 -11.4    & 4.37  &  0.445   & 12.1       \\
   & $0^+_1$      & BEM  &
 -0.205   & 0.347    & -0.038    & 0.441    &  &
  -8.69   & 6.98     & -1.99     & 10.5     \\
\end{tabular}
\end{table}

\begin{table}[t]
\caption{The $0\nu\beta\beta$-decay nuclear matrix elements 
associated with trilinear R-parity violating mode 
for A=76, 82, 100 and 136. The same notation is 
used as in Table I.
}
\label{table.2}
\begin{tabular}{lccccccccccc}
 &  &  & 
\multicolumn{4}{c}{$R_p \hspace{-1em}/\;\:$  SUSY mech.}\\ \cline{4-12} 
 A & f.s. & meth. &
 ${M}_{GT}^{1\pi}$ & ${M}_{T}^{1\pi}$ & ${\cal M}_{}^{1\pi}$ & &
 ${M}_{GT}^{2\pi}$ & ${M}_{T}^{2\pi}$ & ${\cal M}_{}^{2\pi}$ & &
 ${\cal M}_{{\lambda'}_{111}}$ \\ \hline
76 & $0^+_{g.s.}$ &       & 
   1.30  & -1.02 & -24.3 & & -1.34 &  -0.652 & -601.& &    -625. \\
   & $0^+_{1}$    &  RCM & 
  0.254 &  -0.009 & -21.7 & & -0.139 &  -0.014 & -46.3 & &   -68.0 \\
   & $0^+_{1}$    &  BEM & 
  0.482 &  -0.027 & -40.2 & & -0.475 & -0.050 & -158. & &   -198. \\
82 & $0^+_{g.s.}$ &       & 
      1.234 & -0.873 & -31.9 & & -1.258 & -0.572 & -551. & &  -583. \\
   & $0^+_{1}$    &  RCM & 
   0.253 &  0.011 & -23.4 & & -0.135 &  -0.000 & -40.8 & &   -64.2 \\
   & $0^+_{1}$    &  BEM & 
   0.462 &  0.001 & -40.9 & &  -0.449 & -0.030 & -144. & &   -185. \\
100 & $0^+_{g.s.}$ &       & 
   1.433 &  -1.726 & -25.9 & &  -1.525 & -1.048 & -775. & &   -750. \\
   & $0^+_{1}$    &  RCM &
   0.204 &   0.018 & -19.6 & &  -0.102 &  0.017 & -25.6 & &   -45.2 \\
   & $0^+_{1}$    &  BEM & 
   0.532 &  -0.208 & -28.6 & &  -0.509 & -0.129 & -192. & &   -221. \\
136 & $0^+_{g.s.}$ &       & 
   0.606 &  -0.840 & 20.7 & &  -0.742 & -0.543 & -387. & &    -367. \\
   & $0^+_{1}$    &  RCM & 
   0.783 &  0.033 & -72.1 & & -0.383 &   0.021 & -109. & &    -181. \\
   & $0^+_{1}$    &  BEM & 
   0.284 &  -0.084 & -17.7 & &  -0.318 &  -0.076 & -119. & &  -136. \\
\end{tabular}
\end{table}

\begin{table}[t]
\caption{The calculated nuclear matrix elements 
of $0\nu\beta\beta$-decay of $^{76}Ge$
associated with exchange of light and
heavy neutrinos and gluinos for different values of 
$g_{pp}$ within its expected physical range in the RQRPA.
}
\label{table.3}
\begin{tabular}{lccccccccccc}
 & &   \multicolumn{2}{c}{$M^{light}_{<m_\nu >}$}
 & & & \multicolumn{2}{c}{$M^{heavy}_{\eta_{_N}} $}
 & & & \multicolumn{2}{c}{$M^{     }_{{\lambda'}_{111}}$ } \\ 
\cline{2-4} \cline{6-8} \cline{10-12}
$g_{pp}$ & 
$0^+_{g.s.}$ & $0^+_{1}BEM$ & $0^+_{1}RCM$ & &
$0^+_{g.s.}$ & $0^+_{1}BEM$ & $0^+_{1}RCM$ & &
$0^+_{g.s.}$ & $0^+_{1}BEM$ & $0^+_{1}RCM$ \\ \hline
0.8 &   3.8  & 1.34 & 1.30  &
    &  39. & 19.  & 3.7 &
    & -686. & -225. & -71. \\
1.0 &  2.8 & 0.99 & 1.28 &
    & 33. & 16.  & 3.6 &
    & -625. & -198. & -68.\\
1.2 &  1.6 & 0.58 & 1.20 &
    & 27. & 13.  & 3.3 &
    & -564. & -163.  & -63.\\
\end{tabular}
\end{table}

\begin{table}[t]
\caption{The sensitivity factors  $\zeta_{<m_\nu >}$,  
$\zeta_{\eta_{_N}}$,  $\zeta_{{\acute \lambda}_{111}}$ 
[see Eqs. (4.1)] and 
calculated $0\nu\beta\beta$-decay half--lifes $T_{1/2}$  
for transitions to both ground and excited $0^+_1$ states 
of the final nucleus (A=76, 82, 100 and 136) by assuming
$<m_\nu > = 1eV$,  $\eta_{_N} = 10^{-7}$ and 
${\acute{\lambda}}_{111} = 10^{-4}$.
$G_{01}$ is the kinematical factor.
}
\label{table.4}
\begin{tabular}{lcccc}
  & $^{76}Ge$ &$^{82}Se$ & $^{100}Mo$  & $^{136}Xe$ \\ \hline
\multicolumn{5}{c}{ $0^+_{g.s.} \rightarrow 0^+_{g.s.}$ 
$0\nu\beta\beta$-decay transition} \\
 $E_i-E_f$ [MeV]  & 3.067  &   4.027  &   4.055  &   3.503 \\
 $G_{01} $ [$y^{-1}$] & $7.98\times 10^{-15}$ & $3.52\times 10^{-14}$ &
 $5.73\times 10^{-14}$ &  $5.92\times 10^{-14}$ \\
 $\zeta_{<m_\nu >}$ &  2.49 &  4.95 &  7.69 &  1.60 \\
 $\zeta_{\eta_{_N}}$  & 2.90 &  5.64 &  7.10 &  3.43 \\
 $\zeta_{{\acute \lambda}_{111}}$ & 5.57 &  10.9 & 17.9 &  8.92 \\ 
 $T_{1/2}~ (<m_\nu > = 1eV)~[yr]$ &  $4.21\times 10^{24}$ & 
 $1.07\times 10^{24}$ & $4.42\times 10^{23}$ & $ 1.02\times 10^{25}$ \\
 $T_{1/2}~ (\eta_{_N} = 10^{-7})~[yr]$  &  $1.19\times 10^{25}$ &
 $3.14\times 10^{24}$ &  $1.98\times 10^{24}$ & $8.50\times 10^{24}$ \\
 $T_{1/2}~({\acute \lambda}_{111}=10^{-4})~[yr]$ & $1.04\times 10^{25}$ &
 $2.73\times 10^{24}$ &  $1.01\times 10^{24}$ &  $4.07\times 10^{24}$ \\
 & & & & \\
\multicolumn{5}{c}{ $0^+_{g.s.} \rightarrow 0^+_{1}$ 
$0\nu\beta\beta$-decay transition} \\
 $E_i-E_f$ [MeV]  &    1.945  &   2.539 &    2.925 &    1.924 \\
 $G_{01} $ [$y^{-1}$] &  $ 6.58\times 10^{-16}$ & $3.25\times 10^{-15}$ &  
 $1.11\times 10^{-14}$ & $2.81\times 10^{-15}$ \\
\multicolumn{5}{c}{ RCM calculation} \\
 $\zeta_{<m_\nu >}$ &  0.328 & 0.764  & 1.34 & 2.34 \\
 $\zeta_{\eta_{_N}}$   &  0.092 &  0.229 & 0.379 & 0.641 \\
 $\zeta_{{\acute \lambda}_{111}}$   & 0.174 & 0.366 & 0.476 & 0.959 \\
 $T_{1/2}~ (<m_\nu > = 1eV)~[yr]$ &   $2.42\times 10^{26}$ & 
 $4.47\times 10^{25}$ & $1.46\times 10^{25}$ & $4.76\times 10^{24}$ \\
 $T_{1/2}~ (\eta_{_N} = 10^{-7})~[yr]$  & $1.18\times 10^{28}$ & 
 $1.90\times 10^{27}$ & $6.95\times 10^{26}$ &$2.43\times 10^{26}$ \\
 $T_{1/2}~({\acute \lambda}_{111}=10^{-4})~[yr]$ &  $1.06\times 10^{28}$ 
 & $2.42\times 10^{27}$ &  $1.43\times 10^{27}$ & $3.52\times 10^{26}$ \\
\multicolumn{5}{c}{ BEM calculation} \\
 $\zeta_{<m_\nu >}$    &  0.255 & 0.540 & 1.85 & 0.234 \\
 $\zeta_{\eta_{_N}}$   &  0.418 & 0.866 & 1.71 & 0.557 \\
 $\zeta_{{\acute \lambda}_{111}}$    &  0.508 & 1.055 & 2.33 & 0.721 \\
 $T_{1/2}~ (<m_\nu > = 1eV)~[yr]$ &   $4.02\times 10^{26}$ & 
 $8.96\times 10^{25}$ &  $7.59\times 10^{24}$   & $4.77\times 10^{26}$ \\
 $T_{1/2}~ (\eta_{_N} = 10^{-7})~[yr]$  & $5.72\times 10^{26}$ &  
 $1.33\times 10^{26}$ & $3.43\times 10^{25}$ & $ 3.23\times 10^{26}$ \\
 $T_{1/2}~({\acute \lambda}_{111}=10^{-4})~[yr]$ & $1.26\times 10^{27}$ & 
 $2.91\times 10^{26}$ & $5.97\times 10^{25}$ & $6.23\times 10^{26}$ \\
\end{tabular}
\end{table}


\end{document}